# Advances in Microwave Near-Field Imaging: Prototypes, Systems, and Applications

Wenyi Shao and Todd McCollough

Microwave imaging is a science which evolves from detecting techniques to evaluate hidden or embedded objects in a structure or media using electromagnetic (EM) waves in the microwave range (i.e., from 300 MHz to 300 GHz). Microwave imaging is often associated with radar detection such as target detection and tracking, weather pattern recognition, and underground surveillance which are far-field applications. In recent years, due to microwave's characteristic that allows penetration into optically opaque media, short-range applications including medical imaging, nondestructive testing and quality evaluation, through-the-wall imaging, and security screening have been utilized. Microwave near-field imaging occurs when detecting the profile of an object within the short-range: when the distance apart from the sensor to the object is from less than one wavelength to several wavelengths -- This is coarse definition because the size of the antenna and the object is often comparable to the range between them.

A near-field microwave imaging system attempts to reveal the presence of an object and/or an electrical property distribution by measuring the scattered field from many positions surrounding the object. Typically many sensors are placed near the object and either a quantitative or a qualitative algorithm is applied to the collected data. Over the past few decades, both the hardware and software components of a near-field microwave imaging system technology have attracted interest throughout the world. Due to limitations of hardware technology (unavailability of data acquisition apparatus), experimental microwave imaging is very challenging for the pioneers. Examples can be found are the canine kidney imaging experiment carried by Jacobi and Larsen in the 1970s [1]-[3], and active microwave imaging for horse kidney by Jofre and Bolomey [4]. In 1990s, researchers started being able to use microwave signal higher than 1 GHz in real imaging systems. One example is by Bolomey and Pichot, who developed a practical system [5] and designed a planar microwave camera, both operating at 2.45 GHz [6]. However, Probably due to the hardware cost, most of the studies (operating at a few GHz) were still focused on software only. The feasibility of using microwave approaches to image different types of objects have been tested and verified by simulations in a variety of applications. Further, work has been conducted on improving both quantitative and qualitative algorithms to improve simulated reconstruction results. Nowadays, benefitting from the hardware progress and reduction of their cost, researchers are eager to pursue real experimental validations instead of simulations, and, more unique prototypes and commercial systems have been built for various applications. These prototypes and systems are a result of years of dedicated work and it is important to review the advancements in developed prototype systems. The article will provide an overview of the many of the systems designed from different research groups throughout the world, for applications of near-field microwave imaging. The article further outlines challenges faced in current microwave near-field imaging, developmental tendencies of engineers and scientists, and the future outlook.

## Microwave Medical Imaging

Microwave imaging for medical applications has attracted significant interest within the last twenty years. The physics behind microwave medical imaging is that the dielectric properties of abnormal tissue (e.g., malignant tissue) are significantly different than that of normal tissue at microwave frequencies. This difference in dielectric values results in a noticeable contrast in the reconstructed microwave image. Current studies typically focus from 500 MHz to 10 GHz to balance the tradeoff between penetration depth and image resolution.

According to various acquisition setups, the measurement system can be in a passive or active mode. The passive mode systems exploit the principles of radiometry [1]–[10], which produce an image of natural microwave radiation from the human tissue. In active mode systems, microwave radiation is radiated towards tissue and the scattered EM fields are detected and processed. As early as 2000, at Dartmouth College, Meaney's group [11] reported the first clinical prototype for active near-field microwave imaging of the breast. They implemented 16 monopoles in a ring to transmit and receive microwave signals in the 300~1000 MHz range. Tomographic images were created for seven different heights as the ring was moved from the chest wall toward the nipple. This system set the stage for later development in active mode microwave medical imaging. Some prior review papers and books [7]-[16] provide good perspective to microwave imaging technology developed by Meaney and thereafter researchers. Since the purpose of this article is to review hardware microwave imaging systems, our discussion in this section will briefly address some old systems that have been already included in prior review in order to reflect historic information. The focus will be on recent active-mode hardware systems for the early-stage breast cancer detection and brain stroke detection, because which are two prominent medical applications that have been investigated.

*Microwave Breast Imaging*

With cost reduction in RF manufacturing, many microwave breast imaging systems operate at a few Gigahertz to obtain high resolution, where the dielectric property provides a high contrast between malignant and healthy breast tissues (dielectric contrast between malignant and healthy fibroglandular tissue is not high [17]). In 2007, Meaney's group reported an optimized imaging system (as shown in Figure 1a) [23]. In this system, the patient is examined in a prone position. The breast is suspended in a tank of coupling liquid composed of a glycerin and water mixture providing a better impedance match between the skin and a water only coupling liquid used in their earlier system [11]. Sixteen monopole antennas were used to transmit and receive signals at seven frequencies from 500 to 1700 MHz in 200 MHz increments. A multi-static mode system is used, with one antenna in the array serving as a transmitter, while the others are receivers. Data were acquired at seven vertical positions from the chest wall in 0.5 or 1.0 cm increments depending on breast size. Breast phantoms with different heterogeneities were placed in the system and 2-D images were reconstructed with a microwave tomography approach where dielectric values were found. In 2011, they reported their newer result: the system was applied to detect a target inclusion immersed in a plastic breast phantom filled with glycerin and water. The plastic breast-shape phantom, as shown in Fig. 1b and Fig. 1c, was made according to MR scans of a real breast. The permittivity of the target inclusion is 68 at 1300 MHz and the background medium (glycerin and water mixture) is 22.3. 3-D reconstruction of the breast phantom was successfully made for both the permittivity and conductivity as shown in Fig. 1d. The target inclusion can be clearly seen in the reconstructed image.

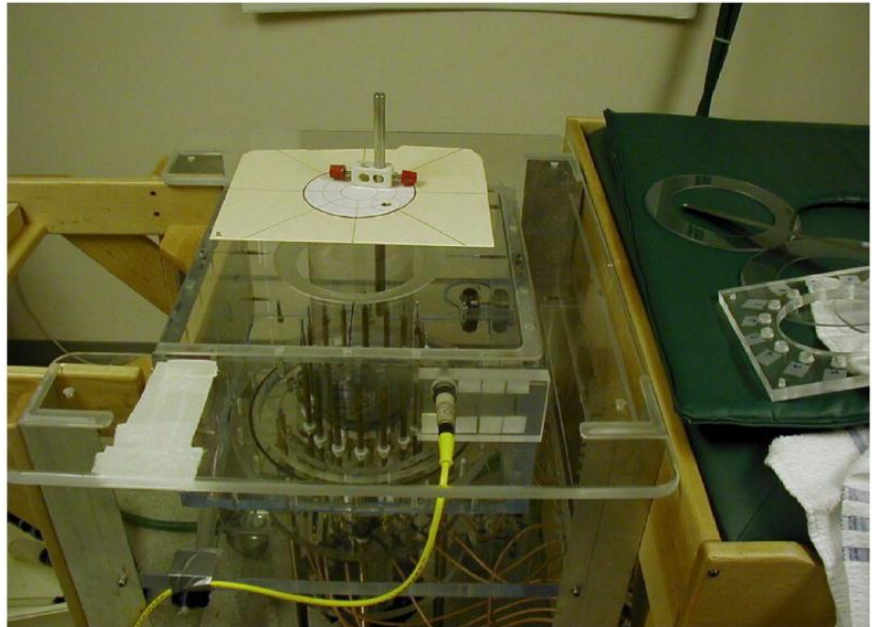

(a)

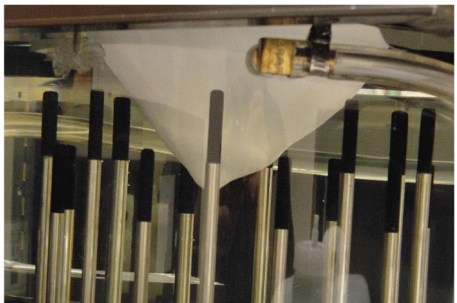

(b)

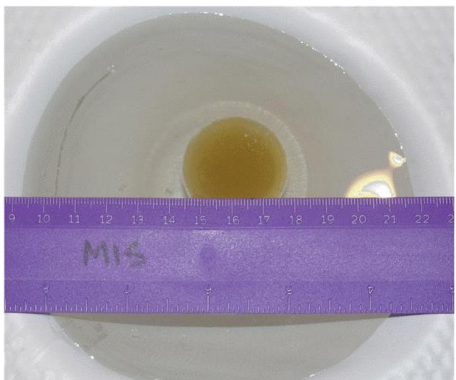

(c)

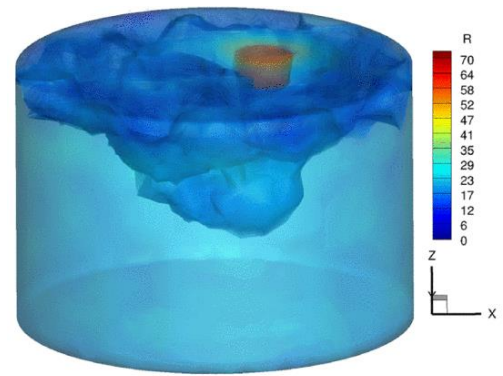

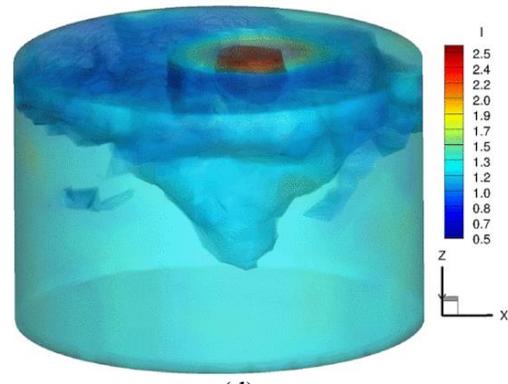

(d)

Figure 1 *(a) Microwave breast imaging system set up at Dartmouth College [23]. Sixteen monopole antennas are immersed in a tank filled with a coupling liquid. (b) side view of the breast phantom. (c) top view of the breast phantom and the target inclusion. (d) reconstructed permittivity (top) and conductivity (bottom).*

A more advanced microwave breast imaging system was developed at the University of Bristol by Craddock and his colleagues [24]-[27], as shown in Figure 2(a) and (b). The breast is placed in a hemispherical cup and a good match to the antennas is provided by using a layer of matching fluid, which may be replaced by a ceramic insert. The first design used 31 ultra-wideband (UWB) slot antennas in the array [24]-[26], and a later design increased the number of antenna elements to 60 [27]. The antenna array was connected to a switching network before being connected to an 8-port vector network analyzer (VNA), allowing for a rapid scan up to 8 GHz. The later system can acquire a total of 1770 unique S-parameter measurements in only 10 seconds. In contrast, the earlier system, with a 2-port VNA, can only acquire a total of 65 S-parameter measurements in 90 seconds. Thus, the later system offered a reduction in the chance of noise due to any slight patient movement. As now the later system has been used for test on 86 patients for clinical trials and achieved a detection rate of 74% (64/86) for wide age range, and achieved 86% in dense breasts []. This result is definitely encouraging because it is comparable or even better than some widely applied clinic methods. Their work opens the pathway from lab prototype study to large scale clinical trials.

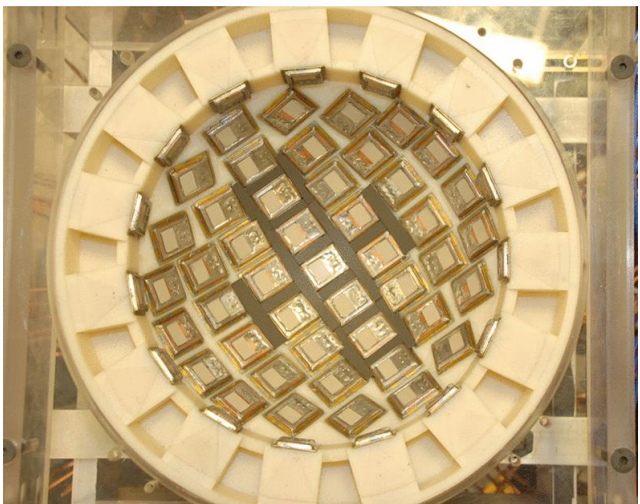
(a)

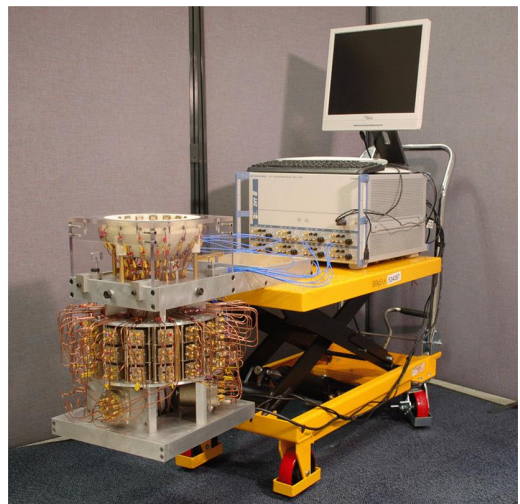
(b)

Figure 2. *UWB breast cancer imaging system developed at the University of Bristol [27]. (a) The UWB conformal antenna array consisting of 60 slot antenna elements. (b) The switching network (below) connects the UWB antenna array (top) and the 8-port VNA.*

At the University of Calgary, Fear and her colleagues developed a prototype system for radar-based breast imaging, with a patient table similar to the one developed at Dartmouth College. A woman lies with one of her breasts suspended through a hole in an examination table. A tank containing immersion liquid, a UWB transceiver antenna [29], and a laser are located under the examination table [30]. The UWB antenna and the laser are mounted on an arm which can rotate about the center of the tank and move in the vertical direction, as shown in Figure 3(c) and (d), to achieve an equivalent effect as using an antenna array. The laser is utilized to first scan the 3-D surface of the breast and define the reconstruction volume. The UWB antenna operating in a monostatic mode sends and receives microwave signals over the frequency range from 2.4 GHz to 15 GHz, but data are recorded from 50 MHz to 15 GHz by the VNA before being converted to the time domain to process [31]. Fear and her group successfully created 3-D images using a confocal imaging algorithm [32] developed by Hagness et al for eight patients with the developed system. Such system has now been upgraded to the $2^{nd}$ generation []. With the assistance of laser, the antenna can be precisely placed in an adaptive matter such that microwaves are normally incident on the skin. Such design may deliver more microwave energy into the breast and then significantly increase the signal to noise ratio. Fear's ground has ever developed another system for breast tissue dielectric permittivity estimation []. Instead of using any coupling liquid, direct contact is made between the breast skin and the sensor array. It will be interesting to see the competition between two mechanisms developed by the same research group.

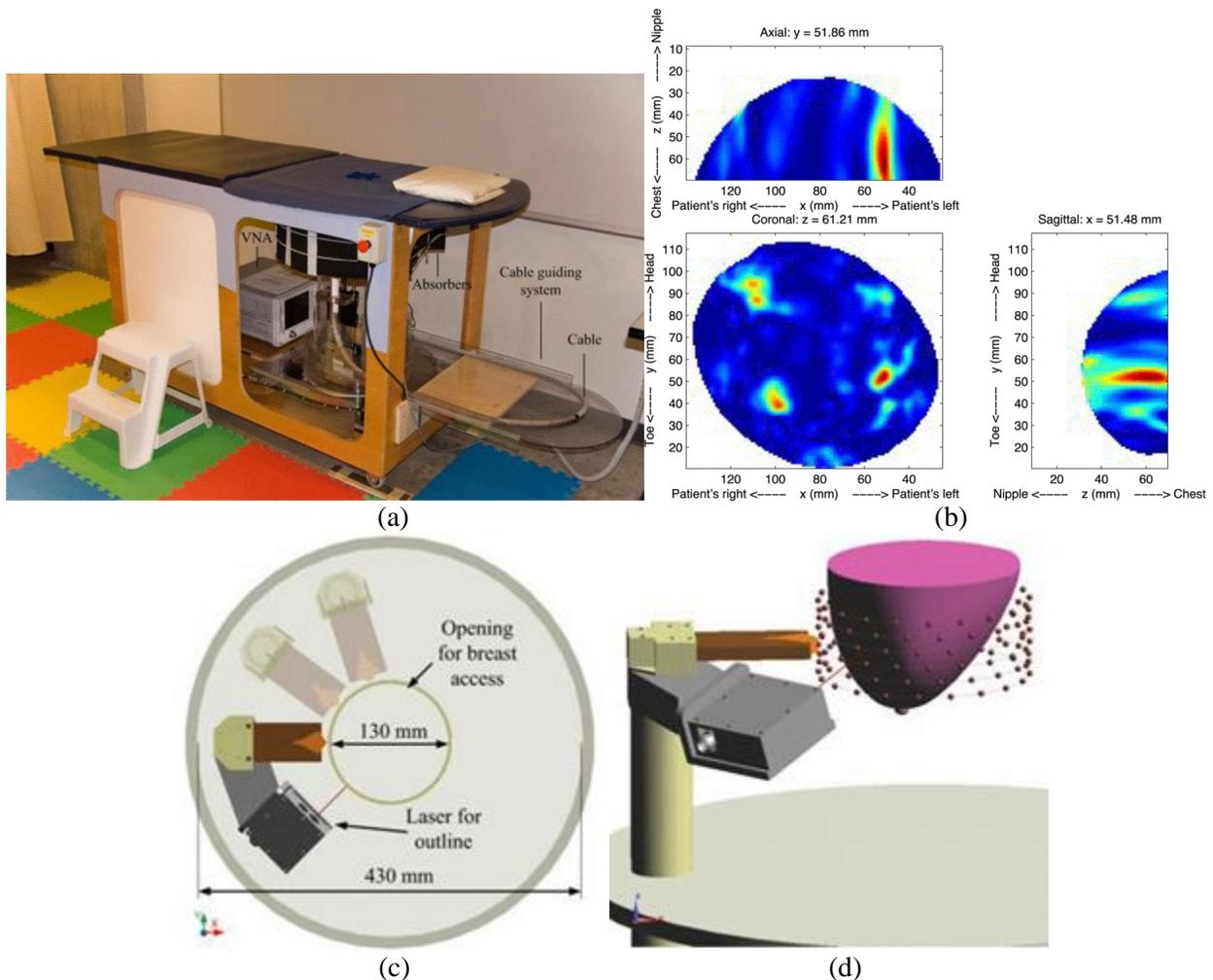

Figure 3. *Monostatic radar-based microwave breast imaging system prototype developed at the University of Calgary [31]. (a) The entire view of the prototype system. (b) Reconstructed images using a confocal imaging algorithm [32] on data collected by the system for a patient. (c) Top view and (d) side view schematics of the antenna and the laser that can move around the breast.*

Using a VNA (especially one with multiple ports) with a microwave imaging system largely increases system cost and size. Persson's research group at Chalmers University of Technology (CUT) developed a compact imaging system for medical diagnostics which does not use a VNA [35]-[37]. Instead, their time domain system contains an impulse generator (to transmit UWB pulses), a high-speed analog-to-digital converter (ADC) (to sample the analog data collected by antenna), and a field-programmable gate array (FPGA) (to store and process digital data and control the entire measurement). In order to solve the problem of insufficient bandwidth of ADC (800 MHz bandwidth), a wideband track-and-hold (T/H) circuit is utilized to track the received signal and freeze the signal ahead of the ADC, and then hold it for a length of time to allow for the ADC to sample. In addition, a direct digital synthesizer (DDS) is used to generate sampling instructions to T/H and the ADC. The signal receiver and the antenna array developed at Chalmers University of Technology are illustrated in Figure 4. An antenna array composed of 20 monopoles were evenly distributed on a circle and no coupling medium was used. A switching matrix was used to select different transmitting and receiving antenna pairs in the array. Finally, a 2-D time-domain inverse algorithm was implemented to reconstruct a dielectric image, illustrated in Figure 4(c), when a cup of vegetable oil was placed in the middle of the circle. Although no clinical test was performed and it was unclear as to what disease the developed system targets, the replacement of a VNA with a customized receiver system is an advance. Such data acquisition unit was also reported by Rubaek et al. [] from the same department at CUT for breast cancer detection. The difference derives from the antenna array which consists of 32 monopoles horizontally positioned in a tank filled with glycerin and water. The monopoles are placed in four rows with eight in each row. In addition, instead of using the FPGA, data are stored in a computer where the signal processing is accomplished with LabView software. Since a SAR aperture is also created in the vertical direction, it is able to create a 3-D image for the object under test, in contrast to reconstructing 2-D image only by Persson.

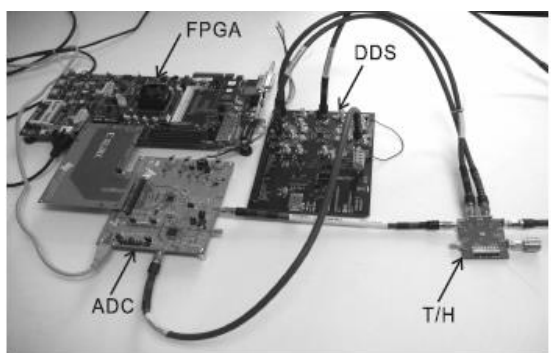 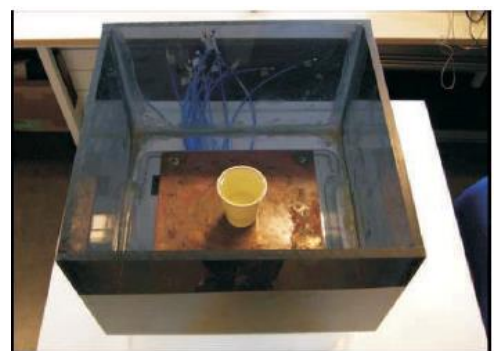

(a)                                    (b)

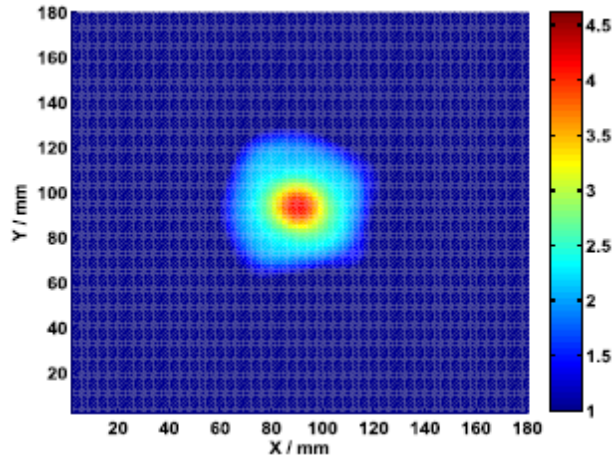

(c)

Figure 4. *Time domain microwave system for medical diagnostics at Chalmers University of Technology [37]. (a) The receiver system. (b) A plastic cup of vegetable oil is placed at the center of the monopole antenna array. (c) A reconstructed image of the cup of vegetable oil.*

Time-domain measurements may offer advantages over frequency-domain measurements, including, faster scan times and more cost-effective measurement devices. Popovic and her team at McGill University also developed time-domain microwave imaging systems for breast imaging [39]-[43]. In the first system developed [39]-[42], 16 antennas were inserted in slots along the outer surface of a hemispherical shaped radome, illustrated in Figure 5(a) and (b), allowing for transmitting and recording both copolarized and cross-polarized pulses using a bandwidth from 2 to 4 GHz. The radome is an alumina bowl in which a breast is placed in. In order to fit women with various breast sizes, the radome was designed to accommodate the largest anticipated size, and an immersion medium was used for smaller breasts to avoid air gaps. Nevertheless, the lossy immersion medium attenuated the responses from the breast, and added additional reflection interfaces between the antenna and breast. Therefore, in a later design, a wearable antenna array consisting of 16 monopoles was embedded into a bra as shown in Figure 5(c) [43]. The antennas are on the inside of the bra and are tangential to the skin surface. Thus, the antennas touch the skin directly and there is no need for an immersion medium. The scan process is a typical multi-static mode in the time domain: a pulse is generated then fed to a switching network which selects the transmitting and receiving antennas, and the received signal is recorded by a picoscope. A complete scan consisting of 16 X 15 signals takes about 6 minutes. The bra-based system was tested on a healthy volunteer over 28 days for breast health monitoring [43]. Consistent imaging results demonstrated data collected by the bra-based system have good repeatability.

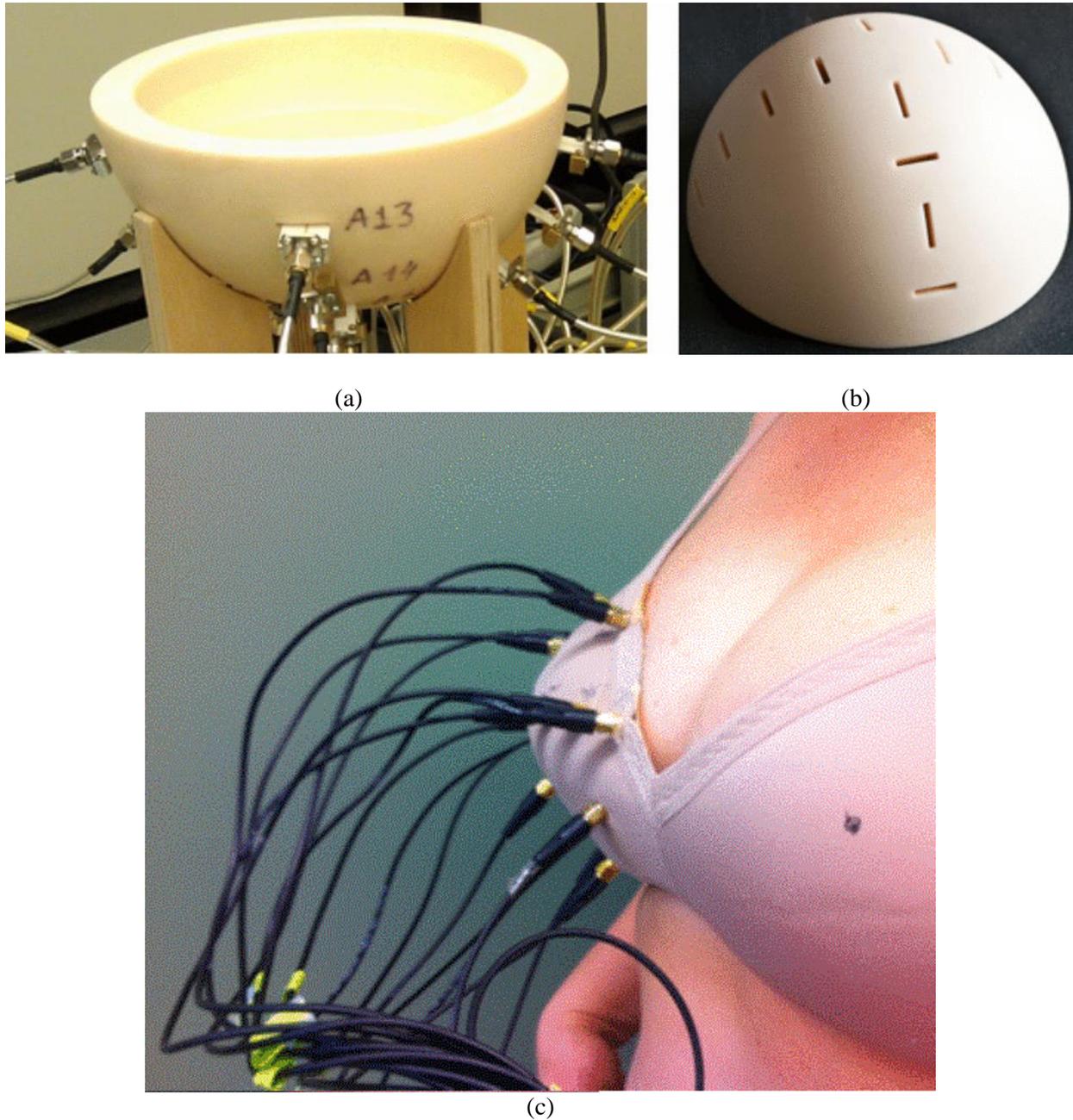

(a)

(b)

(c)

Figure 5. *Time-domain measurement system for breast imaging developed at McGill University. (a) and (b) Views of a radome to house the antennas for a table-based prototype [18]-[21]. (c) A bra-based measurement system [43].*

Another notable example of microwave imaging for breasts was by Vipianna and colleagues at

Politecnico di Torino [44]-[46]. Like the Chalmers University of Technology group, the system

developed uses an embedded platform enhanced with an FPGA to receive and process signals, making the system more practical and able to be commercialized. Additionally, the signal processing speed of FPGA was 20 times faster than on a multicore CPU, allowing for more rapid image reconstructions. Figure 6(a) illustrates the architecture of the prototype system. Breast phantoms were scanned in a coupling liquid (glycerin and water mixture) that the antennas were specifically designed for working in. Only 200-MHz bandwidth between 1.4 and 1.6 GHz were needed to produce a fairly good image due to the use of an interferometric-MUSIC algorithm, which does not require a large bandwidth to detect scatterings inside the breast. Therefore, the system can be produced cost effectively as electronic components in the frequency range of interest are available for low prices.

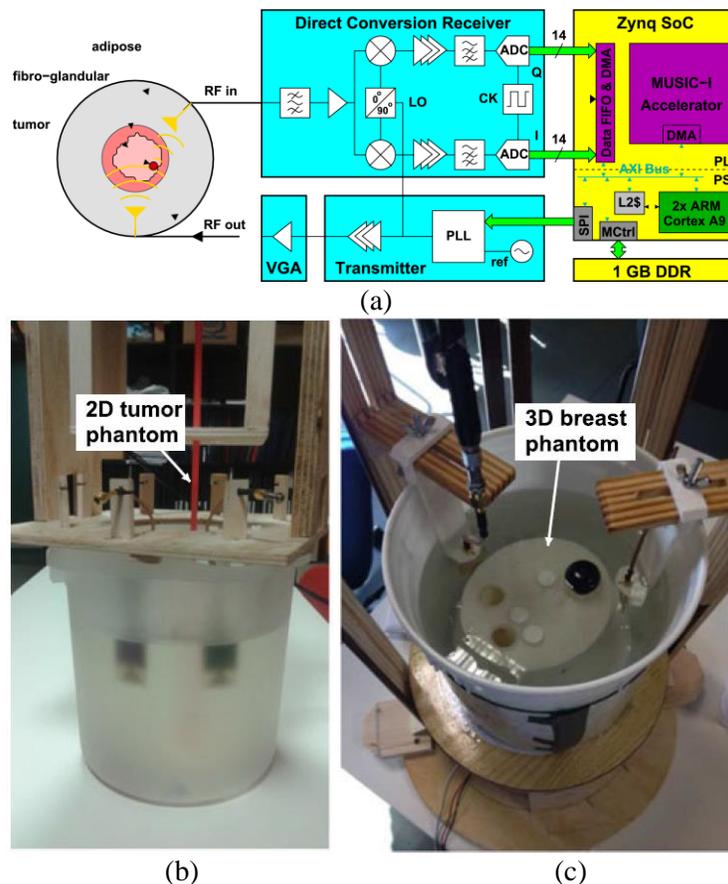

Figure 6. *Microwave imaging breast cancer detection system developed at Politecnico di Torino [46]. (a) Architecture of the prototype system. (b) 2-D trial setup. (c) 3-D trial setup.*

In Japan, Dr. Kikkawa reported a hand-held impulse radar detector for breast tumor detection []. The probe is a 4 × 4 cross-shaped dome antenna array operating in the UWB frequency range from 3.1 to 10.6 GHz, which allows to transmit Gaussian monocycle pulses with width of 160 ps. The detector is designed to be placed on the breast with the patient in a supine position. Fig. X shows the bowl-type probe, the breakdown view of the archetecture for the detector. A step motor is mounted on the top of the detector which can drive the system including the dome antenna array to rotate with 1 degree accuracy. A plastic cover is installed on the antenna dome to protect patient and mitigate frection during rotating. The acquired analog signal is sampled by an ADC with 12 bit accuracy for high resolution. The pusle generator , switching matrix, and sampling module are integrated on the complementray metal-oxide-semicondductor (CMOS) circuit with 65-nm technology. Owing to these modules, the impulse radar imaging system is very compact, compared to all systems that have mentioned above. The detector is connected to a computer by a USB cable for data collection and processing. The confocal algorithm is used for processing the saved data to procude an image.

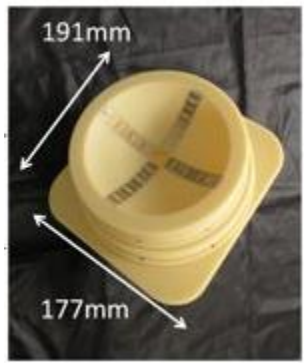

(a)

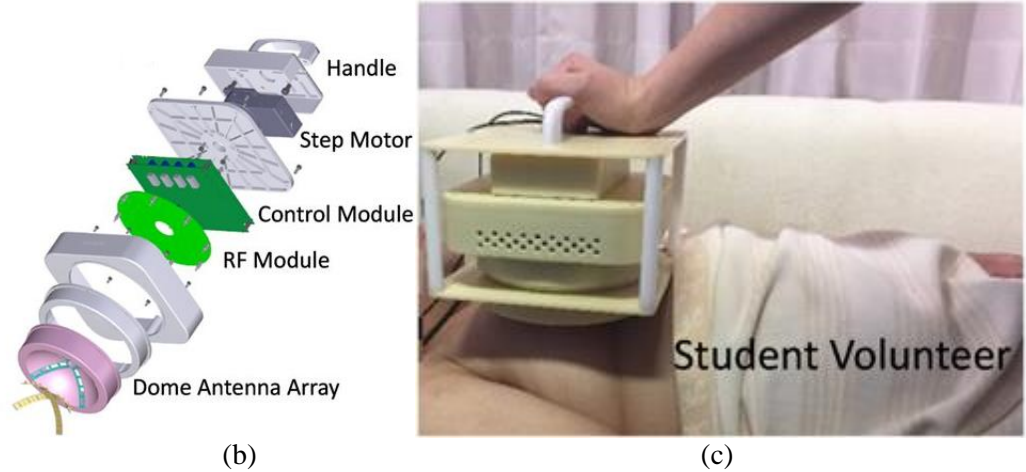

Fig. X. The hand-held breast tumor detector developed at Hiroshima University hospital, Japan []. (a) is the probe which is a dome antenna array. (b) is the breakdown diagram of the system. (c) shows the system testing for a volunteer.

Table 1 shows a comparison of the microwave breast imaging systems described above. The comparison highlight differences between the numbers of antennas used, the frequencies of operation, the hardware used, and the use of a coupling liquid.

Table 1: *Comparison of microwave imaging detection systems based on number of antennas, frequencies of operation, hardware used, and use of a coupling liquid.*

| Group | Antennas | Frequency | Hardware | Coupling Liquid |
|---|---|---|---|---|
| Meaney [6] | Multi-static 16 monopoles | 500 MHz to 1.7 GHz | VNA | Yes |
| Craddock [10] | Multi-static 60 elements | 4 GHz to 8 GHz | VNA | Yes |
| Fear [13] | Mono-static 1 UWB transciever antenna | 2.4 GHz to 15 GHz | VNA | Yes |
| Persson [17] | Multi-static 20 monopoles | 800 MHz to 3.8 GHz | ADC / FPGA | No |
| Popovic [22] | Multi-static 16 monopoles | 2 GHz to 4 GHz | picoscope | No |
| Vipianna [25] | Multi-static 2 wide-band monopoles | 1.4 GHz to 1.6 GHz | FPGA | Yes |

*Microwave Brain Imaging*

A microwave method for human brain imaging has not yet attracted quite as much interest as microwave breast imaging. This technology is based upon studies that tissue malignancies, blood supply, hypoxia, acute ischemia, and chronic infarction significantly change dielectric properties of the affected tissues at microwave frequencies [48]. By exposing head tissues to a low level of microwave energy and capturing the scattered signal by an antenna, estimation of dielectric profiles of brain tissues can be made.

Pioneering research for microwave brain imaging can be traced back to 1982 by Lin and Clark's work [49], in which the detection of cerebral oedema (accumulation of water in the brain) was experimentally tested by using a 2.4 GHz microwave signal through a simple head phantom. Modern active mode microwave imaging for a brain or head scan was first published in 2008, with Semnov's study at Keele University [50] and Persson's research group at Chalmers University of Technology [51]. Semnov simulated microwave measurements with the number of transmitters and receivers being 32 X 32 or 64 X 64 (multi-static mode) for a 2-D model of a head with a radius of 11 cm, containing a few soft tissues, skull, and a small stroke area. Then data was processed in an image reconstruction procedure using the Newton approach [52] over a frequency range from 0.5 GHz to 2.5 GHz. A "stroke-like" area with a radius of 2 cm was seen in the image reconstructed at 1 GHz frequency. Semnov suggested against use above 1 GHz for microwave brain tomography because of high microwave energy attenuation in the brain. In Persson's study, a low cost patch antenna with a triangular shape was designed to be of small size and lightweight, which can potentially serve as the antenna element in an array for brain stroke monitoring. To improve impedance matching, the use of a matching liquid with relative permittivity of 78 was used, either placed in a bag between the antenna and the head (for the high frequency case), or with the antenna immersed in (for the low frequency case). Finally, a system containing 8 such antennas with various distances from each other was designed and simulated for overall performance characterization. A later development involved a fabricated array on a helmet the head is placed in [53], and the elements in the array increased to 10 or 12, illustrated in Figure 7(a) and (b), respectively.

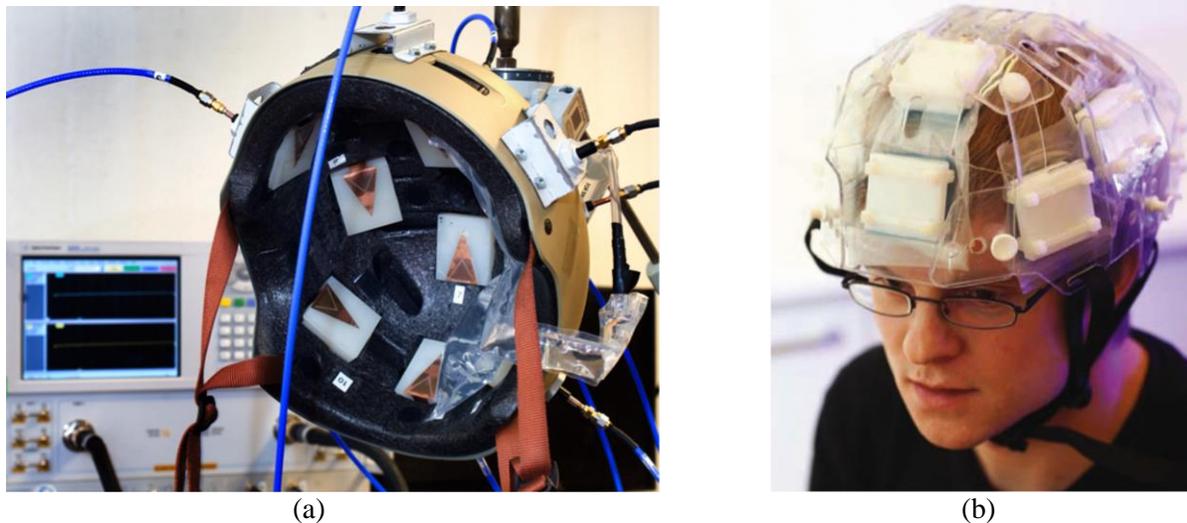

(a)            (b)

Figure 7. *Antenna array developed for brain stroke detection at Chalmers University of Technology [53]. (a) Ten patch antennas are mounted on a helmet with plastic bags used for matching liquid. (b) Twelve patch antennas are mounted on a custom-built supporting structure.*

At the University of Queensland in Australia, brain imaging for stroke injuries detection has been intensively studied by Abbosh and his group [54]-[58]. The first detection system [54], [56] developed by the group was in 2013, as shown in Figure 8(b). Sixteen corrugated tapered slot antennas (operating frequency from 1 to 4 GHz) are fixed on a table, and are equally distributed around a head phantom with a fixed distance of 5 mm from the head boundary. The antennas work in a monostatic mode. To increase the number of signals to be processed in image reconstruction without increasing the number of antennas so as to avoid unwanted mutual coupling between them, the table holding the antennas can rotate around to collect data from different angular positions. To carry out a practical brain scan measurement, a realistic 3-D head phantom was built as shown in Figure 8(a) consisting of a suitable mixture of water, corn flour, gelatin, agar, sodium azide, and propylene glycol so as to mimic different tissues of the brain with realistic dielectric properties. The dielectric properties of the fabricated tissues agree with measurements with less than 3% error across the frequency band from 1 to 4 GHz. An ellipsoid object with dielectric properties equivalent to blood was emulated as the stroke region and inserted inside the head phantom. S-parameters were collected by using a VNA and algorithms were used to reconstruct 2-D images. Stroke regions were successfully localized with high contrast. Due to high loss in the head at high frequencies, in later work, antenna elements have been replaced by compact unidirectional antennas with an operating frequency from 1.1 to 2.2 GHz [58], which provides a compromise between penetration depth and image resolution. The second system developed by Abbosh's group, uses only one unidirectional antenna (covering a frequency band from 1.1-3.4 GHz) to send and receive signals to and from the head phantom, placed on a rotatable support as shown in Figure 8(c). A portable custom-made microwave transceiver, Agilent N7081A, replaces a VNA to send and receive signals. N7081A is a high-speed, small, low-cost and modular wideband transceiver device that can operate over a wide bandwidth from 0.1 MHz to 4 GHz and offers a maximum dynamic range of 80 dB. It is controlled by an in-home operation system (IHOS) installed on a PC via USB or LAN connections for post data processing. The IHOS manages the entire detection process with a flowchart depicted in Figure 8(d).

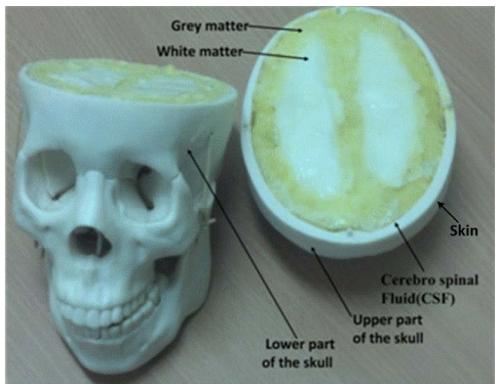

(a)

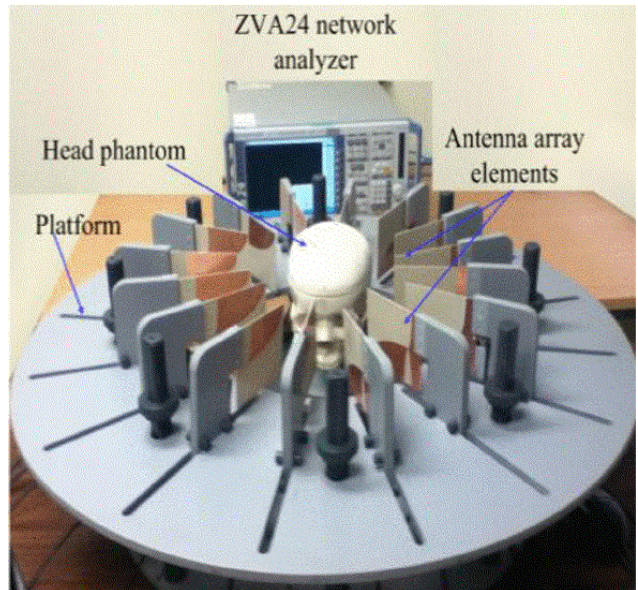

(b)

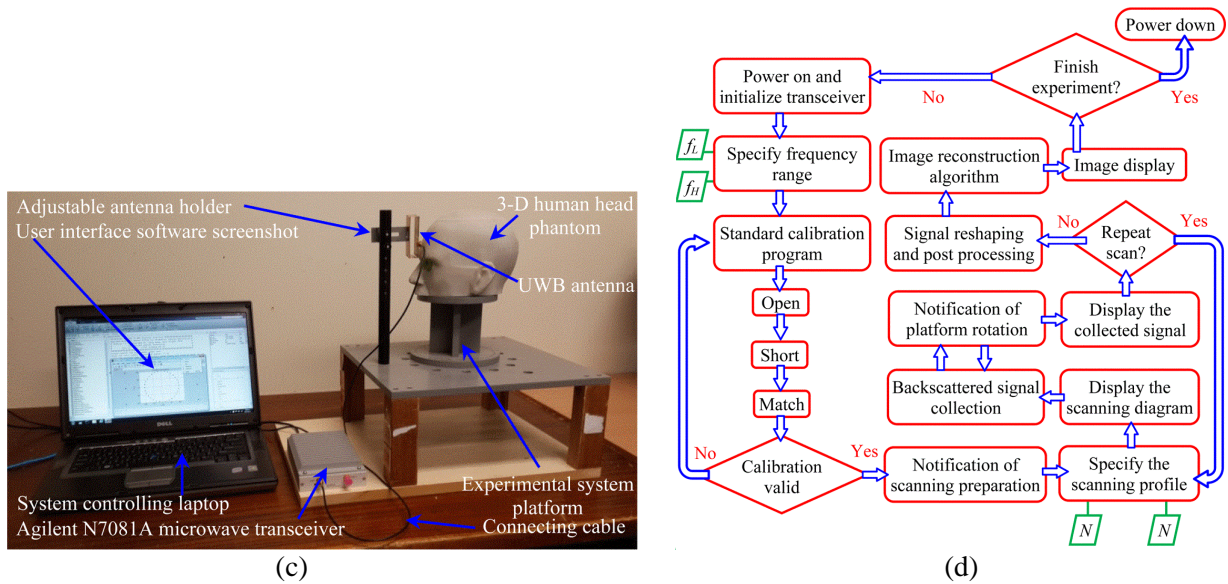

Figure. 8. *Head phantom and monostatic imaging system for brain stroke detection developed at the University of Queensland. (a) Head phantom. (b) Antenna array consisting of 16 corrugated tapered slot antennas on a rotatable platform [54][56]. (c) The second system composed of one fixed antenna and the phantom on a rotatable support [57]. (d) A graphical flowchart of IHOS.*

A brain stroke imaging prototype system has been developed by EMTensor GmbH, a company founded by Dr. Semenov, along with Prof. Pichot and colleagues at the Universite Cote d'Azur in France [59], [60]. The system consists of a cylindrical metallic chamber composed of 5 rings of 32 transmitting and receiving antennas as shown in Figure 9(a). The antennas are ceramic loaded open-ended waveguides, operating from 0.9 to 1.8 GHz. The system operates in a multistatic mode using a switching matrix utilized to connect the antennas and a network analyzer, which results in a 160 X 160 matrix of S parameters. The data acquisition cycle of the system is fully electronically controlled, allowing for a total data acquisition to occur in around 30 seconds. As illustrated in Figure 9(b) and (c), the chamber is in a horizontal position allowing for easy positioning of a human head within the imaging zone. A special thin membrane is used for isolating the human head from a matching liquid and keeping the liquid within the chamber. Two measurements, one with an empty chamber, and another with the head, are conducted to later obtain the scattered field of the head by a subtraction. The collected data is wirelessly transferred to a remote computing center for high-performance post processing. Images for a numerical head model have been reconstructed for verification of the developed prototype. An experimental implementation and verification has not yet been seen.

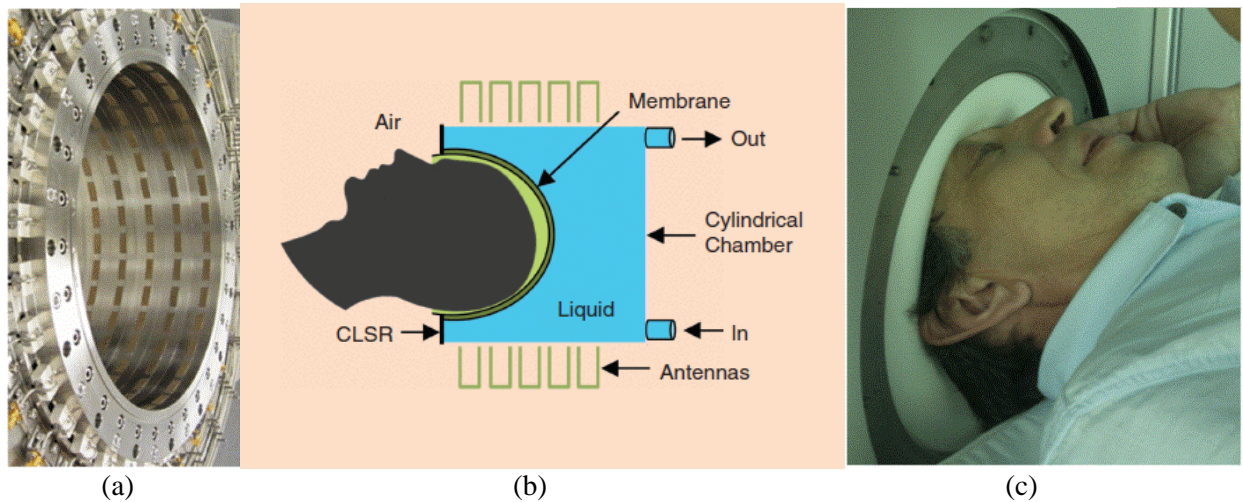

Figure 9. *Brain stoke imaging prototype system developed by EMTensor GmbH  [59], [60]. (a) Chamber with 160 antennas in 5 rings. (b) Architecture of the chamber. (c) A human head measurement with the prototype system.*

### Other Medical Diagnostic Applications

Besides breast cancer and brain stroke imaging, the near-field microwave approach has also been applied in other medical diagnoses such as knee injury, skin cancer detection, and lung cancer detection.

The knee is one of the most frequently injured joints of the human body, with s traumatic injuries in the young and tissue degeneration in the old. Fear et al. utilized a simulation method and some simple experiments to test the feasibility of using radar-based methods to detect meniscal tears [61], [62], tendon [62], [63], and ligaments [63] at microwave frequencies. However, no prototype systems have been reported by her group for diagnosis for knee pathologies.

For skin cancer detection, some early studies utilized microwave reflectometry to differentiate cancerous skin tissues from healthy skin tissues using an open-ended coaxial probe with a VNA [64] [65]. The reflection coefficient was compared between cancerous skin tissues and adjacent normal skin tissues from 300 MHz up to 6 GHz. Recent studies are more interested in exploring the properties of cancerous skin tissue at millimeter wave (mm-wave) range because of better resolution and no deep penetration requirements for skin measurement. In 2013, Taeb et al. proposed a reflectometry device operating at 42 and 70 GHz for non-invasive early stage skin cancer detection [66]. Clinical trials tests were performed on a number of volunteer patients who suffered from basal cell carcinoma. The device presented by Taeb provides a low-cost solution for fast and accurate skin cancer detection and has the potential to be used for fast tissue inspection during surgery. Most of the open-ended rectangular waveguide probes implemented for skin detection have footprints in the range of several square millimeters and their EM field penetrates deep into the subcutaneous fat, i.e., they measure the average of a larger volume compared to the size of an early stage (very small) skin tumors. In fact, malignant melanoma grows from the bottom of the epidermis, typically starting at a depth of around 100 $\mu$m and first infiltrates the epidermis. Therefore, to achieve an accurate measurement, a high-quality mm-wave probe with sub-millimeter sensing depth and high lateral resolution is desired. In 2015, Topfer et al. invented a broadband probe operating from 90 to 104 GHz [67], which consists of a dielectric-rod waveguide that is metallized and tapered towards the tip to achieve high resolution by concentrating the electric field in a small sample area. The sensing depth was from 0.3 to 0.4 mm, which is adapted for detecting early-stage skin tumors before metastasizing. The lateral resolution can be as high as 0.2 mm, allowing for resolving small skin tumors and even the inhomogeneities within a tumor. Other than skin cancer detection, in 2017, Gao et al. used mm-wave reflectometry approach to accurately assess the degree of a burn on human skin [68]. In the test, mm-wave reflectometry and imaging were verified as capable of distinguishing between healthy and burned skin as the dielectric properties of the two are significantly different in a 26.5 – 40 GHz frequency band.

Other than the knee diagnosis and skin cancer detection, we noticed the recent reports regarding using microwave imaging for lung cancer detection, which is based on the hypothesis that significant difference exists between the dielectric properties of cancerous lung tissue and healthy lung tissues at microwave frequency band. In Australia, Abbosh's group applied a very similar lab system developed for brain stroke detection (Fig. 8b) to perform lung cancer microwave imaging experiments []-[].The head phantom was replaced by an artificial torso phantom comprised of ribs, muscles, skin, heart, lungs and abdomen block. To mimic the cancerous lung tissue, an artificial cancer of dimension $1 \times 1 \times 2$ $cm^3$, made of a mixture of water, corn flour, gelatin, agar, sodium azide, propylene glycol and NaCl salt, was inserted inside the torso phantom. The mixture has the same dielectric properties of lung cancer across the working bandwidth from 1.5-3 GHz. Instead of rotating the table that supports the antenna like in Fig. 8b, only one antenna was adopted with position fixed and the phantom was rotated in every $30°$ to achieve 12 mono-static data. And then a frequency-domain algorithm was applied to the acquired data to obtain an image. In fact, the challenge of imaging the torso is the same to the head microwave imaging. It is difficult to balance the penetrating-in-body microwave energy and the imaging resolution. From the reconstructed image that Abbosh's group obtained, the dimension of the "cancerous region" is much bigger than the artificial cancer that was inserted in the torso phantom. Thus, good ideas are expected before any breakthrough can be seen in the torso and head microwave imaging.

## Nondestructive Testing

Nondestructive testing (NDT) technologies are often needed to determine a component of an object or to quantitatively measure characteristics. Traditional NDT uses ultrasound or X-ray. Due to good penetration capability and cost efficiency, microwave imaging techniques have been explored in NDT. Known applications of microwave NDT, see for example Figure 10, include moisture measurements, wall thickness measurements, measurements of paint thickness on carbon composites, quality control, e.g. presence of seams in composite materials, measurements of material permittivity, corrosion and precursor pitting detection in painted aluminum and steel substrates, flaw detection in spray-on foam insulation, and the acreage heat tiles of the Space Shuttle. In recognizing growing importance of NDT, an expert committee for microwave and Terahertz (THz) procedures of the German Society of NDT and the microwave testing committee of the American Society for NDT (ASNT) were founded in 2011 and 2014, respectively. Standardization work has begun along with prototype systems in both reflection mode and through transmission mode.

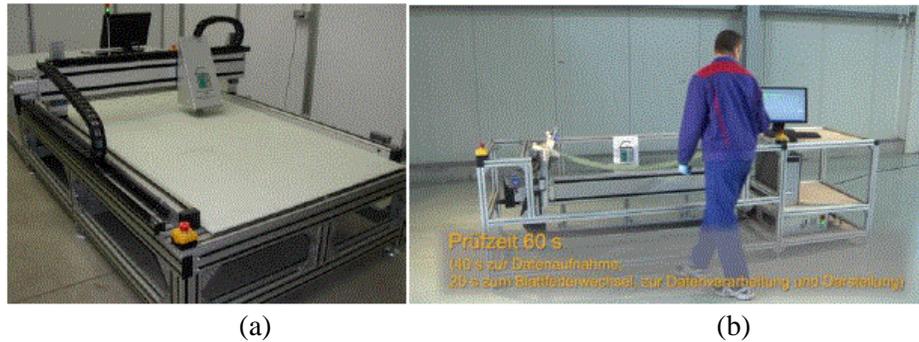

(a)                    (b)

Figure 10. *Stationary microwave test systems [71] for (a) planar substrates and (b) glass-fiber reinforced plastic leaf springs for automobiles.*

Zoughi's research group at the Missouri University of Science and Technology [72] utilized microwave imaging for NDT to investigate a mortar specimen with four different embedded rebars located at a depth of 2 cm below the surface of the mortar as shown in Figure 11(a). A linear scan was performed above the surface of mortar denoted by the red arrow in Figure 11(a) and (b). The linear scan, was carried out by an antenna located 13.8 cm from the surface of the mortar, and spanned a 23-cm range along the $x$ direction with a step size of 0.1 cm. Reflection coefficients were measured and recorded by a VNA from 8.2 to 12.4 GHz. A piecewise-SAR algorithm and a Wiener filter-based layered SAR algorithm allowed for an image successfully specifying the rebars' position, as shown in Figure 11(c).

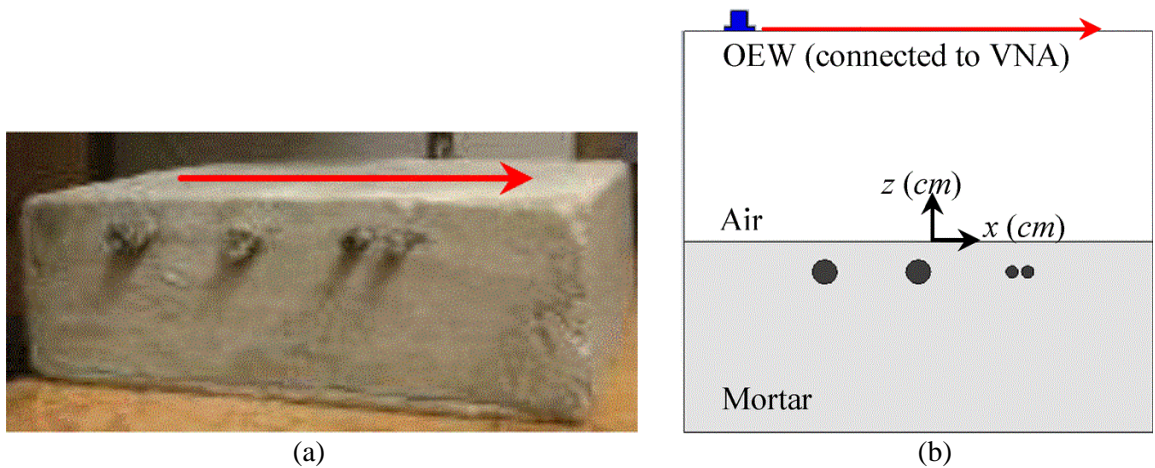

(a)                    (b)

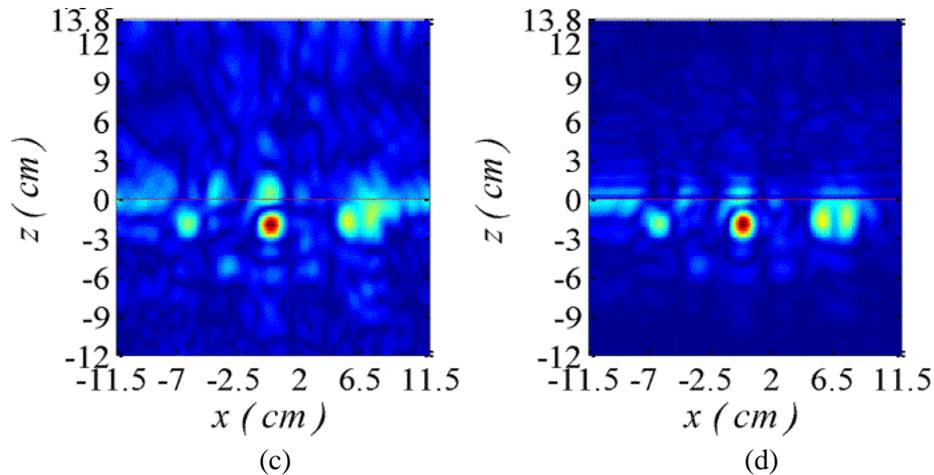

(c)                                             (d)

Figure 11. *Microwave nondestructive detection experiment conducted the Missouri University of Science and Technology [72]. (a) A mortar specimen with four rebars under test. (b) Schematic diagram for the experiment setup: a probe connected to a VNA moves along the surface of mortar specimen to collect reflection coefficients. (c) Reconstructed image by a piecewise-SAR algorithm. (d) Reconstructed image by a Wiener filter-based layered SAR algorithm.*

More interesting, a wideband microwave camera for real-time 3-D imaging still developed by Zoughi's group was reported in 2017 []. The camera operating in the 20-30 GHz frequency range can produce real -time 3-D images, causing the potential to provide real-time inspection and diagnosis in the nondestructive testing, biomedical and security applications. A monostatic antenna array composed of 16 1-D array, with each consisting of 16 elements, is built in a printed circuit board to transmit and receive signal (Fig. X). The entire system, including the antenna array, the switch drivers, analog multiplexer, and other support circuitry such as power regulators and power sequencing circuitry for power amplifier are assembled up with a total size of 26 cm × 21 cm × 18 cm (Fig. X). An awesome video showing the working performance of the built system was uploaded by the team and can be found by https://youtu.be/RE-PPXmtTeA.

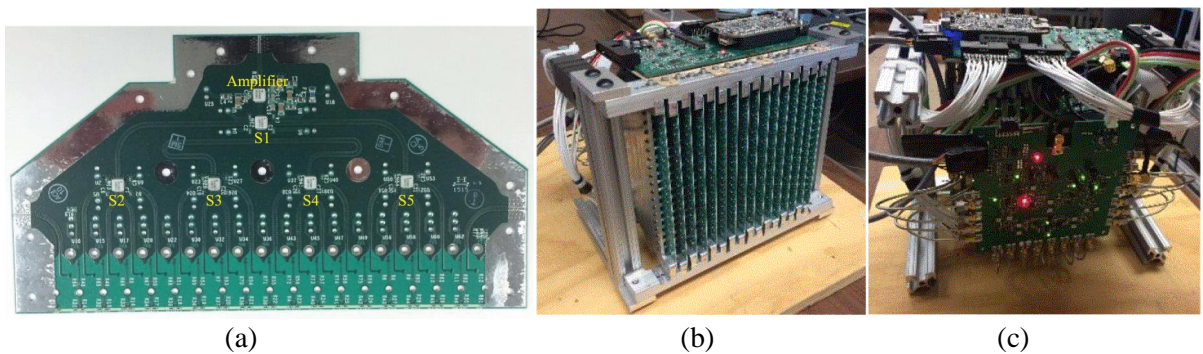

(a)                              (b)                              (c)

Fig. X. (a) The 1-D antenna array, (b) front side showing the array aperture, and (c) back side showing the source board and the switching/control circuit.

Researchers formerly with the University of Michigan and now with the University of Southern California in Moghaddam's group created a microwave imaging system to image changes in the dielectric constant due to the application of microwave heating [74], [75] as shown in Figure 12. Microwave imaging of differential temperature is based on that dielectric properties of water change as a function of temperature. They used 36 bow-tie patch antennas designed to operate at 915 MHz with a 2-port VNA in an isopropyl alcohol and water coupling medium. They imaged a 4 cm ping-pong ball as it cools from 55 to 22 degrees Celsius designed to emulate a thermal therapy focal spot size in their imaging cavity.

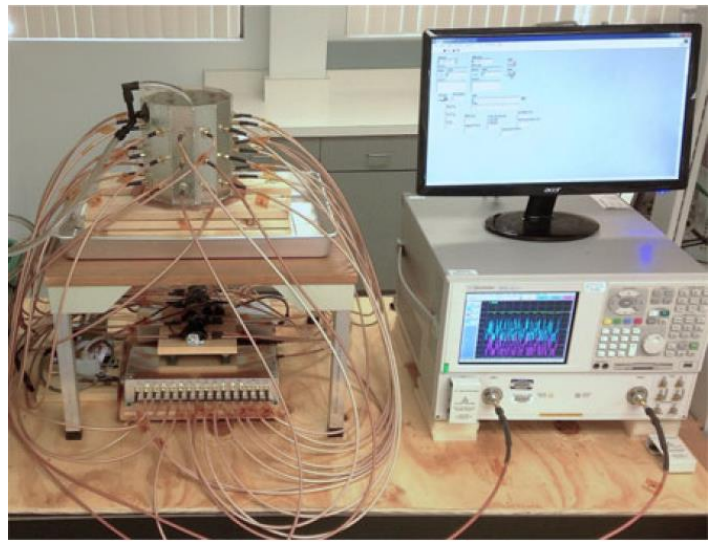

Figure 12. *Microwave imaging differential temperature monitoring measurement system by the University of Southern California [75].*

Collaborators from the University of Applied Sciences of Southern Switzerland, Centro di Senologia della Svizzera, and the University of Genoa including Pastorino's research group developed a multi-static mode prototype system first used for the inspection of dielectric materials including wood objects [76], [77]. In contrast to Moghaddam's design that uses an array and a number of cables, a pair of antennas is used in conjunction with a rotating table that an object is placed on. In a later development, plastic extension arms were connected to the antenna arms to allow them to be immersed in a coupling medium [78]. Measurements were collected using a VNA. A breast phantom placed in vegetable oil was placed in the system and measurements were collected every 22.5 degrees from 2 to 10 GHz with a 200 MHz increment. Figure 13(a), (b), and (c) shows the experimental setup, breast phantom, and reconstructed image using the inversion algorithm in [79].

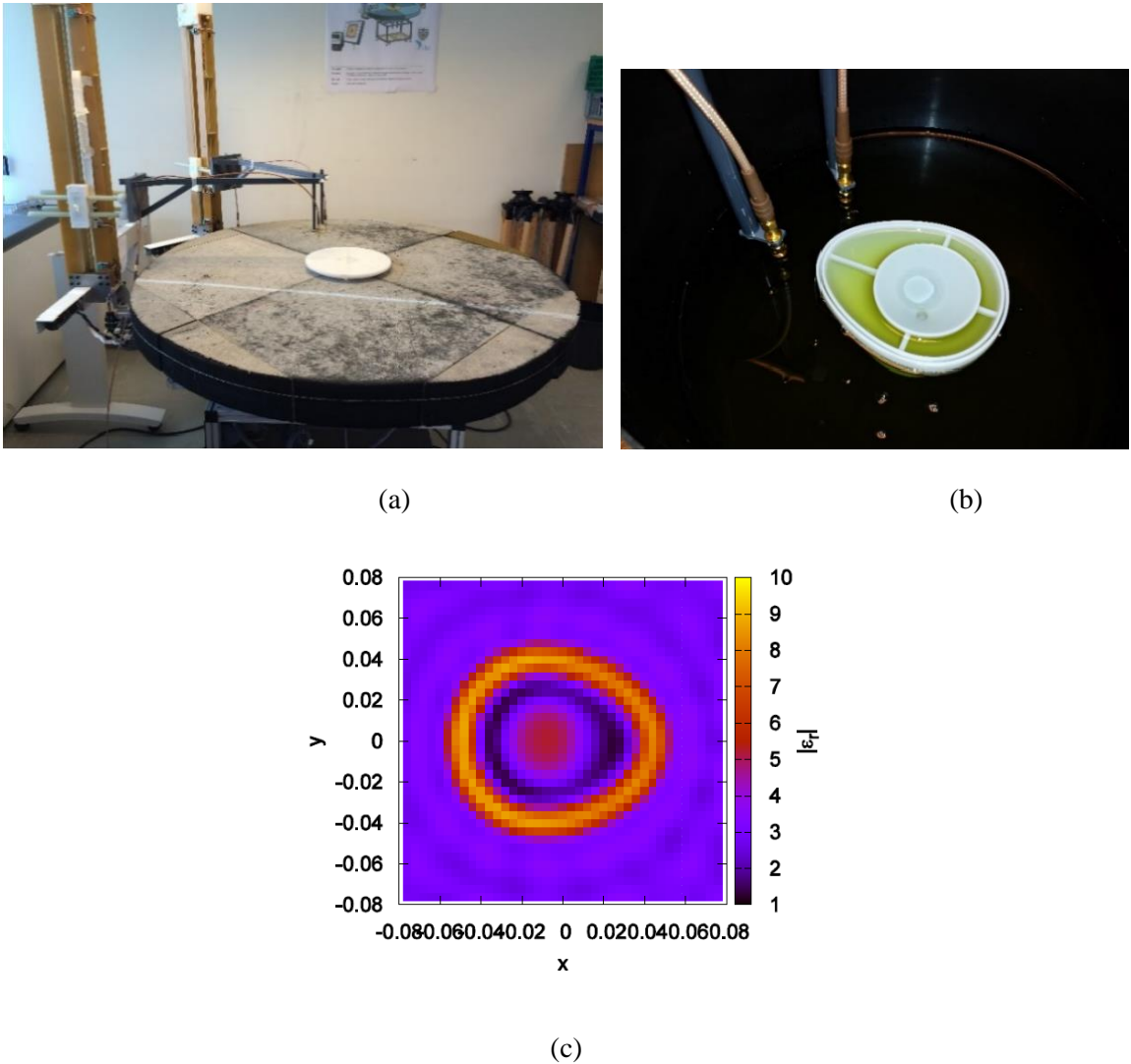

(a)                                                        (b)

(c)

Figure 13. *Microwave measurement system by Collaborators from the University of Applied Sciences of Southern Switzerland, Centro di Senologia della Svizzera, and University of Genoa [78]. (a) Setup for frequency-domain measurement. (b) Breast phantom. (c) The reconstructed image of the breast phantom object with the inversion algorithm of [79].*

A prototype system developed by Ellumen, Inc. [80]-[82] uses both reflected and transmitted signals in a multi-static mode for microwave NDT. Similar to Pastorino's system, a pair of antennas were used but are fully electronically controlled to rotate about an object from 0 to 360 degrees with one-degree accuracy with one antenna serving as the transmitter and the other as the receiver. A wooden circular tray is placed at the center to support an object and can move in the vertical direction. Microwave measurements can be conducted in the frequency domain when antennas are connected to a VNA or in the time domain when the transmitter antenna is connected to a signal generator and the receiver antenna is connected to an oscilloscope as shown in Figure 14(a) and (b). There are advantages of using movable antennas instead of an array containing many elements. This includes avoiding potential unwanted mutual couplings between elements in the array, no need for a costly and bulky switching matrix network, and to allow for more advanced high-gain antennas to be adopted beyond just a small and simple antenna such as monopole or dipole. A complete multi-static scan for both antennas transmitting and receiving from a 0 to 360 degrees takes a few minutes, depending on the number of signals and the length of each signal. A phase confocal imaging algorithm [83] is used to produce an image of the cylindrical object as shown in Figure 14(c), in which the hollow structure is successfully revealed. The software to control the system consists of a series of steps as shown in Figure 14(d). The software controls the antennas to move to any position on the rails by setting the starting and ending positions and the angle increments, along with allowing for multiple measurements taken at each location and the average saved to achieve a lower-noise contained signal.

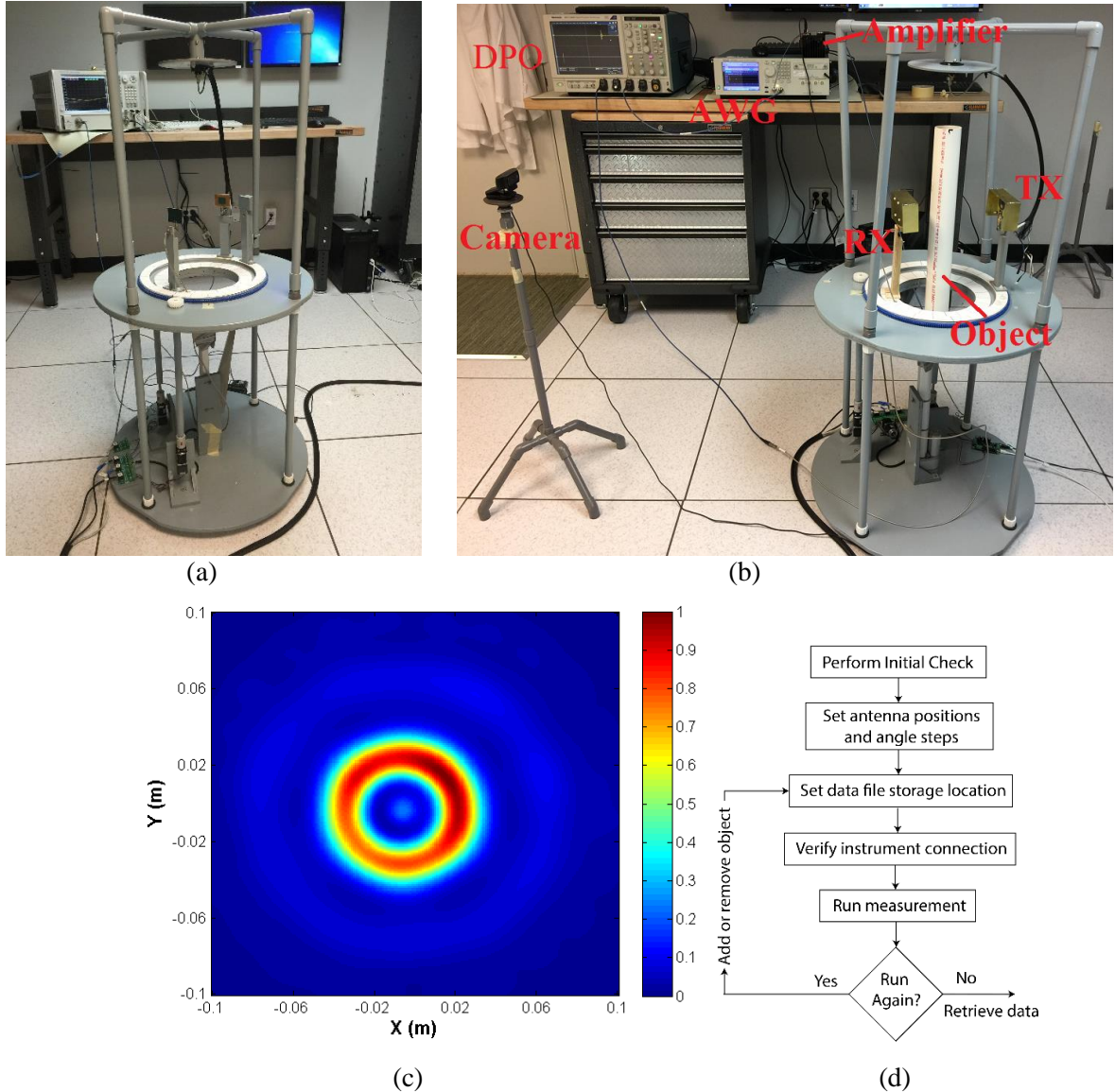

Figure 14. *Microwave measurement system by Ellumen Inc. with antenna movement controllable in a transmit-receive mode. (a) Setup for frequency-domain measurement [80]. (b) Setup for time-domain measurement [81]. (c) The reconstructed image of the cylindrical object with a phase confocal imaging algorithm [83] using the data collected by the system in [80]-[82] (d) Diagram of the important operation steps of the software.*

Researchers at the University of Manitoba in LoVetri's group have developed a microwave imaging approach using a resonant air-filled metallic chamber [84] building upon some of their earlier work [85]. Unlike the Pastorino and Ellumen systems, no rotatable antennas or rotatable object is present. They used printed reconfigurable monopole antennas that can be turned on or off using PIN diodes. They also differed from some other systems described by using a metallic chamber to allow better control over unwanted reflections, potentially allowing the incorporation of a lossless or low-loss matching medium and to shield what is in the chamber from external electromagnetic noise. The system was used to image a wooden cube and nylon cylinder as shown in Figure 15(a) and (b). Reconstructed images, as in Figure 15(c), successfully showed the targets at 1.75 GHz.

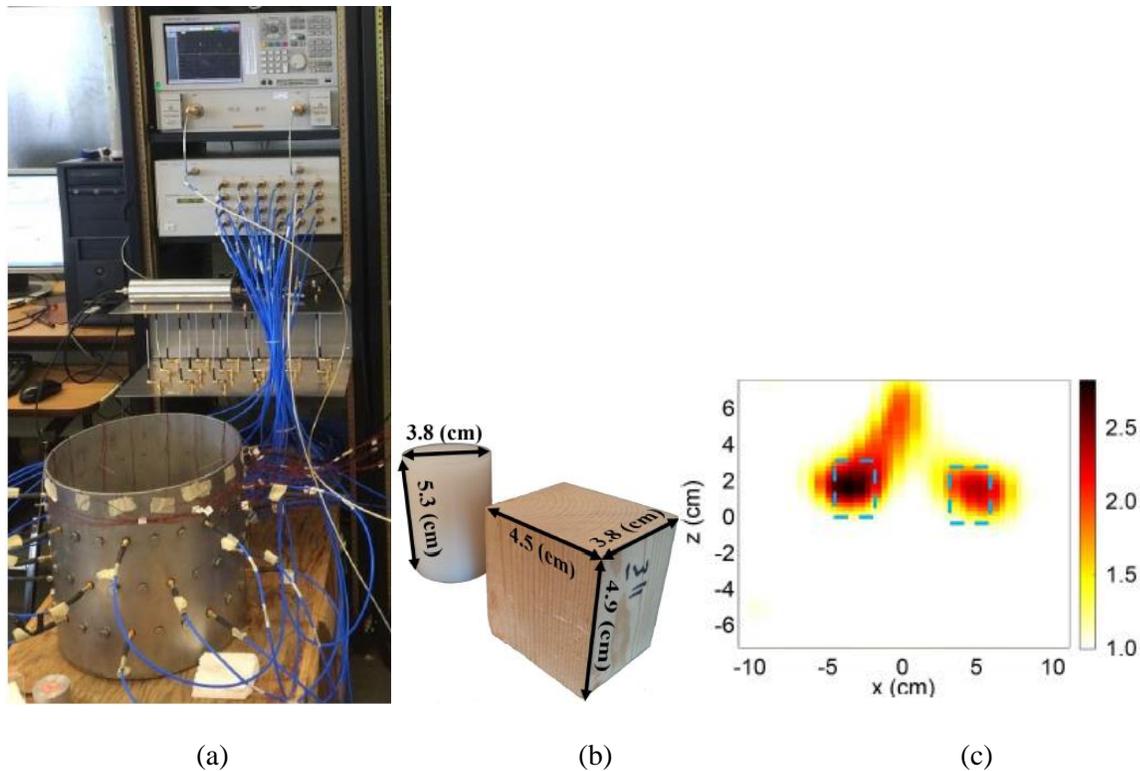

(a)                                    (b)                                    (c)

Figure 15. *Microwave measurement system by the University of Manitoba [85]. (a) Setup for frequency-domain measurement. (b) Wooden cube and nylon cylinder. (c) The reconstructed image of the wooden cube and nylon cylinder at 1.75 GHz.*

## Through-The-Wall Imaging

Through-the-wall microwave imaging is one of the important emerging microwave techniques in recent years. Such an imaging system provides enhanced situational awareness in a variety of civilian and military applications. Such systems not only detect the presence of the targets behind walls, but also provide information concerning each target's location, motion, size, and backscattering cross section (RCS). Compared to traditional radar applications, through-the-wall radar imaging faces challenges such as unknown dielectric characteristics and thickness of the wall, signal attenuation in the wall, and unknown target movement.

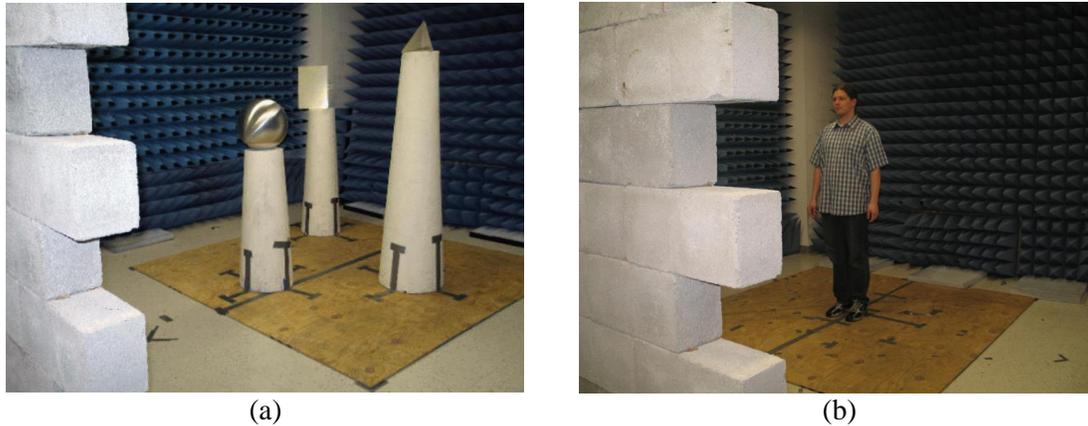

(a)                                         (b)

Figure 16. *Tests for through-the-wall radar imaging by a collaborated research group from AGT Group GmbH, Technische Universitat Darmstadt, and Villanova University [88]. (a) A few objects with different shapes located behind a wall. (b) A human located behind the wall.*

Many approaches have been attempted in testing through-the-wall radar imaging techniques. The most widely used approach is the wideband or UWB radar mechanism, which contains an antenna array, or one antenna measuring at multiple locations to form a synthetic aperture [86]-[92]. The bandwidth covered is typically from a few hundred MHz to several GHz. The backscattered signals measured by all antenna elements in the array (or one antenna at all locations) are accounted for their time delaying and weighting, and are finally synthesized to produce an image. An example of this is shown in Figure 17. Another example is a device developed by Akduman's group [] which uses a pair of horn antenna operating between 0.8-5 GHz. The antenna pair works in a bistatic mode and moves together along the wall to collect the signal from different positions. The experiment was performed in a big anechoic chamber as well. Image of the objects behind the wall was reconstructed by processing the acquired data based on an inverse scattering algorithm (microwave tomography). A second approach used is by using a frequency-modulated continuous wave (FMCW) radar [94], [95] or step-frequency continuous wave (SFCW) radar [96]. This is based on the detection of a Doppler frequency shift due to the target's motion, such as heartbeat and breath. In addition, other approaches including the time reversal technique [97] and the tomographic inverse scattering approach [98] have been used in through-the-wall microwave imaging. The through-wall imaging has achieved commercial success. An interesting video available at https://walabot.com/diy shows a portal microwave device that can be connected to a cell phone is able to detect objects behind the wall and the motion of a mouse inside the wall.

## Security Screening

The microwave imaging technique has been proposed to detect concealed weapons at major transportation hubs to replace the conventional X-ray screening method, as it may cause less potential health risks of X-ray, especially for pregnant women and infants. The non-ionizing characteristic, and the low cost of the microwave imaging system makes it more likely to be adopted in security screening.

Microwave imaging for security applications uses a high frequency which is reflected by human skin yet still penetrates through most fabrics. Thus, a 3-D shape reconstruction of the human along with any metal they may have on them is possible as microwave cannot penetrate through metals. Likely due to hardware restrictions, there are only a limited number of studies in this area. In 2001, an engineering team at Pacific Northwest National Laboratory (Sheen et al.) reported the first 3-D whole-body imaging using the principles of microwave holography []. The goal was to achieve a real-time imaging (on the order of 3 to 10 s) for concealed weapon detection. A prototype imaging system utilizing a 27-33 GHz linear sequentially switched array was developed and tested for scanning concealed explosive. Later they developed the microwave holographic imaging technology for multiple applications with different frequency range [], for example, 3-D imaging of a helicopter using an impulse radar with nominal 1-5 GHz, X-band (8-12GHz) imaging of Bradley fighting vehicle, 3-D GPR imaging with twin waste storage tanks operating at 200-400 MHz, and 40-60 GHz test and 10-20 GHz test for detecting concealed weapons attached to a mannequin.

In 2011, Zhuge, et al. [101], [102] attempted to use the UWB method with a multi-input multi-output (MIMO) array to deliver high-resolution 3-D images for concealed weapon detection in quasi-real time. This system successfully combines UWB, MIMO, and SAR technology. The MIMO array consists of four antipodal Vivaldi antennas as transmitters and eight such antennas as receivers, operating at a center frequency of 11.15 GHz and 150% fractional bandwidth (2.8-19.5 GHz). Elements in the MIMO array are connected to a network analyzer through a multiport switch, and the entire array is mounted on a computer-controlled mechanical scanner allowing for both azimuth and elevation scan to form a synthetic aperture for a 3-D volumetric imaging, as shown in Figure 17(b). When a weapon attached to a 1.8-m-high-and-0.5-m-wide mannequin (covered with an aluminum foil to mimic a shield for the weapon) is exposed to the MIMO array, a 3-D image of the mannequin can be achieved by processing the received data. It is possible to determine the shape of the weapon that was attached to the mannequin as shown in Figure 17(c).

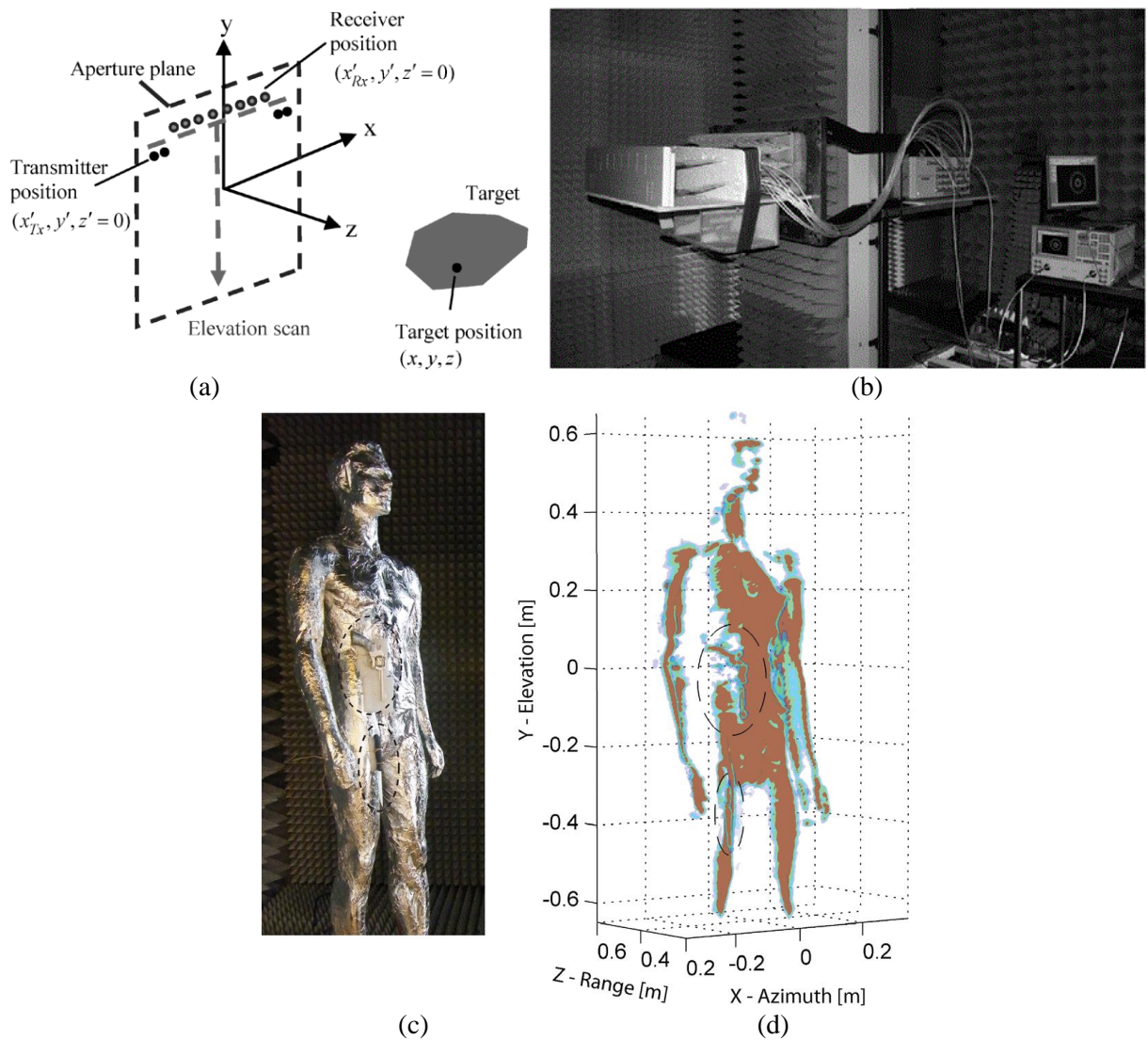

(a)

(b)

(c)

(d)

Figure 17. *UWB MIMO-SAR measurement system setup with network analyzer in an anechoic chamber at Delft University of Technology [101]. (a) Schematic diagram of the MIMO array structure and its movement for a scan. (b) MIMO-SAR configuration with additional multiport switch. (c) A gun and a knife attached to a mannequin covered with an aluminum foil and (d) its 3-D imaging result with the developed system.*

Due to technical progress in RF manufacturing and reduction in costs, mm wave [103], [104] and THz [105] imaging have been considered in security applications. The usage of a shorter wavelength leads to better range resolution. In [103], an active mm-wave SFCW radar operating at 59-61 GHz was tested to image an aluminum foil wrapped toy gun covered with a cloth piece. In the experiment, a 2-D scan composed of 30 X 18 locations with 0.02 m spacing in both the vertical and the horizontal directions were performed. Imaging results validated the idea of using mm wave imaging radar system for concealed weapon detection. In contrast, THz imaging uses even higher frequencies to achieve higher spatial resolution. In [105], Knipper et al. compared the effect of using 0.35 and 0.85 THz for a passive security camera to detect a weapon under clothes. THz imaging was found to be influenced by signal absorption in fabric, especially wet clothes, causing a 0.85 THz band to be a less desirable option than 0.35 THz, although 0.85 THz provided a better resolution.

## Outlook

Over the past few decades, microwave near-field imaging has experienced conceptualizations, theoretical analyses, simulation tests, and experimental validations. Due to technical developments both in the hardware manufacturing and the software, many prototype systems for a variety of applications have emerged since the early 2000s. Microwave scientists and engineers are working towards optimizing current prototypes in order to bring these systems into production and more widespread use.

One trend for current microwave near-field imaging systems is that bulky and expensive signal recording tools, such as a VNA and an oscilloscope, are replaced by more compact, more cost-effective, and customized instruments such as a high-speed ADC accompanied with an FPGA. This trend will allow for current prototypes to lead the way towards the first commercial microwave imaging product. In addition, the growth of this technology will further extend from research institutions and academia into industrial research and development. It can be expected that in the not too distant future, more applications of microwave imaging techniques and their products will emerge and lead to benefits to humans and society.

This paper has been accepted by IEEE Microwave Magazine and published on a future issue.